\documentclass[english,conference]{IEEEtran}
\usepackage[T1]{fontenc}
\usepackage[latin9]{inputenc}
\usepackage{babel}
\usepackage{amsmath}
\usepackage{amsthm}
\usepackage{amssymb}
\usepackage{stmaryrd}
\usepackage[unicode=true,
 bookmarks=true,bookmarksnumbered=false,bookmarksopen=false,
 breaklinks=false,pdfborder={0 0 1},backref=false,colorlinks=false]
 {hyperref}
\hypersetup{pdftitle={Line Spectral Estimation from Partial Observations},
 pdfauthor={Maxime Ferreira Da Costa and Wei Dai},
 pdfsubject={ISIT 2017 - Line Spectral Estimation from Partial Observations},
 pdfkeywords={Line spectral estimation; Atomic norms; Dimensionality reduction; Super-resolution}}

\makeatletter

\newcommand{\noun}[1]{\textsc{#1}}

\theoremstyle{plain}
\newtheorem{thm}{\protect\theoremname}
\theoremstyle{plain}
\newtheorem{lem}[thm]{\protect\lemmaname}
\theoremstyle{remark}
\newtheorem{rem}[thm]{\protect\remarkname}
\theoremstyle{plain}
\newtheorem{cor}[thm]{\protect\corollaryname}
\theoremstyle{plain}
\newtheorem{prop}[thm]{\protect\propositionname}

\makeatother

\providecommand{\corollaryname}{Corollary}
\providecommand{\lemmaname}{Lemma}
\providecommand{\propositionname}{Proposition}
\providecommand{\remarkname}{Remark}
\providecommand{\theoremname}{Theorem}

\begin{document}

\title{Low Dimensional Atomic Norm Representations\\
in Line Spectral Estimation}

\author{Maxime Ferreira Da Costa and Wei Dai\\
\IEEEauthorblockA{Department of Electrical and Electronic Engineering, Imperial College
London, United Kingdom\\
Email: \{maxime.ferreira, wei.dai1\}@imperial.ac.uk}}
\maketitle
\begin{abstract}
The line spectral estimation problem consists in recovering the frequencies
of a complex valued time signal that is assumed to be sparse in the
spectral domain from its discrete observations. Unlike the gridding
required by the classical compressed sensing framework, line spectral
estimation reconstructs signals whose spectral supports lie \emph{continuously
}in the Fourier domain. If recent advances have shown that atomic
norm relaxation produces highly robust estimates in this context,
the computational cost of this approach remains, however, the major
flaw for its application to practical systems.

In this work, we aim to bridge the complexity issue by studying the
atomic norm minimization problem from low dimensional projection of
the signal samples. We derive conditions on the sub-sampling matrix
under which the partial atomic norm can be expressed by a low-dimensional
semidefinite program. Moreover, we illustrate the tightness of this
relaxation by showing that it is possible to recover the original
signal in \emph{poly-logarithmic time} for two specific sub-sampling
patterns.
\end{abstract}

\global\long\def\trans{\mathsf{T}}

\global\long\def\herm{*}

\global\long\def\splus{\mathrm{{\scriptscriptstyle +}}}

\global\long\def\linmeasure{\mathcal{L}}

\global\long\def\toep{\mathcal{T}}

\global\long\def\rmop{\mathcal{R}_{M}}

\global\long\def\trace{{\rm tr}}

\section{Introduction\label{sec:Introduction}}

\subsection{Background on line spectral estimation}

The line spectral estimation problem aims to recover the frequencies
of a complex time signal that is assumed to be sparse in the spectral
domain from its discrete measurements $x\in\mathbb{C}^{n}$, uniformly
acquired at a sampling frequency $F_{s}$. More precisely, we suppose
that the sampled time signal is supported over a small number of frequencies
$s^{\star}$ and that we dispose of observations $x\in\mathbb{C}^{n}$
of the form

\begin{equation}
\forall j\in\left\llbracket 0,n-1\right\rrbracket ,\quad x_{j}=\sum_{k=1}^{s^{\star}}c_{k}^{\star}e^{i2\pi f_{k}^{\star}j},\label{eq:SpikesModel}
\end{equation}
whereby $\left\{ f_{k}^{\star}\right\} _{1\leq k\leq s^{\star}}$
is the ordered set containing the $s^{\star}$ spectral components
generating the signal $x$, and $\left\{ c_{k}^{\star}\right\} _{1\leq k\leq s^{\star}}$
the one of their associated complex amplitudes. Both of those sets,
as well as their cardinality $s^{\star}$, are supposed to be unknown.
We highlight at this point that the particularly of this model is
that the frequencies $\left\{ f_{k}^{\star}\right\} $ can be drawn
\emph{continuously} in $\left[0,1\right)$ and \emph{are not constrained
to belong to some finite discrete grid, }unlike in the classic compressed
sensing framework. The ground truth spectral distribution of $x$,
denoted $\hat{x}^{\star}$, is therefore constituted of $s^{\star}$
Dirac spikes located at the frequencies $\left\{ f_{k}^{\star}\right\} $
forming the spectral support of $\hat{x}^{\star}$.

It is not difficult to the see that this problem is ill-posed, in
the sense that there are infinitely many estimators of the ground
truth spectral distribution $\hat{x}^{\star}$ that are consistent
with the measurement vector $x$. Among all of those estimators, the
one considered to be optimal in the line spectral estimation framework
is the one returning the consistent spectral distribution $\hat{x}_{0}$
having the sparsest possible spectral support. Alternatively, this
estimator can be formulated as the output of the \emph{non-convex}
minimization program
\begin{equation}
\hat{x}_{0}=\arg\min_{\hat{x}\in D_{1}}\left\Vert \hat{x}\right\Vert _{0},\;\text{subject to }x=\mathcal{F}_{n}\left(\hat{x}\right),\label{eq:L0-FullObservation}
\end{equation}
whereby $D_{1}$ denotes the set of tempered integrable spectral distributions,
$\mathcal{F}_{n}:D_{1}\rightarrow\mathbb{C}^{n}$ is the inverse discrete-time
Fourier operator, and $\left\Vert \cdot\right\Vert _{0}$ is the form
mapping $D_{1}$ to $\left[0,\infty\right]$ counting the (potentially
infinite) cardinality of the spectral support.

The study of line spectral recovery under the paradigm of convex relaxations
has been gaining in popularity after the tightness results of such
approaches were demonstrated in the pioneer works \cite{Candes2014a,Candes2012,Tang2013}.
It has been shown in \cite{Tang2015} that a separation criterion
on the spectral support of $\hat{x}^{\star}$ of the from $\Delta_{\mathbb{T}}\left(\hat{x}^{\star}\right)=\min_{i\neq j}\left\{ \left|f_{i}^{\star}-f_{j}^{\star}\right|\right\} \geq\frac{C_{{\rm nec}}}{n-1}$
is always necessary to guarantee the tightness of convex approaches
for a constant $C_{{\rm nec}}=\frac{1}{\pi}$. On the other hand,
it was shown in \cite{Candes2014a} and enhanced in \cite{Fernandez-granda2015}
that $C_{{\rm suf}}=2.56$ is sufficient to ensure the recoverability
of $\hat{x}_{0}$ from a convex surrogate. Since then, many direct
extensions of Program (\ref{eq:L0-FullObservation}) were proposed.
Among them, we mention an extension to multi-dimensional spectra \cite{Chi2015,Yang2015},
to multiple measurement vectors \cite{Li2014a}, and to the spectral
blind deconvolution framework \cite{Yang2016}. Line spectrum estimation
theory can be viewed as a particular case of the spikes deconvolution
problem presented in \cite{Duval2015}.

\subsection{Notations}

The adjunction of ${\bf X}$ is denoted ${\bf X}^{\herm}$, whenever
${\bf X}$ is a vector, a matrix, or a linear operator. The transposition
of ${\bf X}$ is written ${\bf X}^{\trans}$. The set of complex square
matrices of dimension $n$ is denoted ${\rm M}_{n}\left(\mathbb{C}\right)$
and vectors of $\mathbb{C}^{n}$ are indexed in $\left\llbracket 0,n-1\right\rrbracket $
so that every vector $u\in\mathbb{C}^{n}$ writes $u=\left[u_{0},\dots,u_{n-1}\right]^{\trans}$.
The trace operator is denoted $\trace\left(\cdot\right)$. We define
by $\toep_{n}:\,\mathbb{C}^{n}\rightarrow{\rm M}_{n}\left(\mathbb{C}\right)$
the Hermitian Toeplitz generator in dimension $n$, such that for
all $u\in\mathbb{C}^{n}$ , $\toep_{n}\left(u\right)$ is the Hermitian
Toeplitz matrix whose first row is equal to $u$. Its adjoint $\mathcal{T}_{n}^{*}$
is characterized for every matrix $H\in\mathrm{M}_{n}\left(\mathbb{C}\right)$
by
\[
\forall k\in\left\llbracket 0,n-1\right\rrbracket ,\quad\mathcal{T}_{n}^{*}\left(H\right)\left[k\right]=\left\langle \Theta_{k},H\right\rangle =\trace\left(\Theta_{k}^{\herm}H\right),
\]
whereby $\Theta_{k}$ is the elementary Toeplitz matrix equals to
$1$ on the $k^{\textrm{th}}$ upper diagonal and zero elsewhere.
Moreover, for every matrix $M\in\mathbb{C}^{m\times n}$, $m\leq n$,
we denote by $\mathcal{R}_{M}$ the operator given by
\begin{align*}
\mathcal{R}_{M}:\;\mathcal{\mathbb{C}}^{m} & \rightarrow\mathrm{M}_{m}\left(\mathbb{C}\right)\\
v & \mapsto\mathcal{R}_{M}\left(v\right)=M\mathcal{T}_{n}\left(M^{\herm}v\right)M^{\herm}.
\end{align*}
Its adjoint $\mathcal{R}_{M}^{*}$ is consequently characterized for
every matrix $S\in\mathrm{M}_{m}\left(\mathbb{C}\right)$ by $\mathcal{R}_{M}^{*}\left(S\right)=M\mathcal{T}_{n}^{*}\left(M^{\herm}SM\right).$

\subsection{Atomic norm minimization}

Atomic norms were analyzed in \cite{Chandrasekaran2012} as a generic
way to regularize sparse inverse problems defined over continuous
dictionaries. The underlying idea consists in considering the dictionary
of interest $\mathcal{A}$ as a set of building blocks called ``atoms'',
and to endow the search space $E$ with the norm induced with the
gauge of $\mathcal{A}$ defined by
\begin{equation}
\forall x\in E,\;\left\Vert x\right\Vert _{\mathcal{A}}=\inf_{t>0}\left\{ x\in t\,{\rm conv}\left(\mathcal{A}\right)\right\} .\label{eq:AtomicNormGauge}
\end{equation}
The atomic ball defined by $\left\{ x\in E:\;\left\Vert x\right\Vert _{\mathcal{A}}\leq1\right\} $
is by construction the smallest convex set containing the dictionary
$\mathcal{A}$, and one may expect, by analogy with $\ell_{1}$ minimization
in the discrete compressed sensing framework, that atomic norm minimization
has a high sparsity promoting power.

Atomic norms have been introduced in the context of line spectral
estimation in \cite{Tang2013,Bhaskar2013}. Detailed performance guarantees
of this use can be found in \cite{Tang2013a}. In this context, the
set of underlying atoms for Model (\ref{eq:SpikesModel}) takes the
form $\mathcal{A}=\left\{ a\left(f,\phi\right),\;f\in\left[0,1\right),\phi\in\left[0,2\pi\right)\right\} $,
where each atom $a\left(f,\phi\right)\in\mathbb{C}^{n}$ writes for
every $f\in\left[0,1\right)$ and $\phi\in\left[0,2\pi\right)$

\begin{align*}
a\left(f,\phi\right) & =\left[e^{i\phi},e^{i\left(2\pi f+\phi\right)},\cdots,e^{i\left(2\pi\left(n-1\right)f+\phi\right)}\right]^{\trans}\\
 & =e^{i\phi}a\left(f,0\right).
\end{align*}
Using Carathéodory's theorem on convex hulls, any vector of ${\rm conv}\left(\mathcal{A}\right)$
can be expressed by a convex combination of at most $n+1$ points
in $\mathcal{A}$ and the atomic norm reformulates
\begin{align}
\left\Vert x\right\Vert _{\mathcal{A}} & =\inf_{\substack{c_{k}\geq0,\;\phi_{k}\in\left[0,2\pi\right)\\
f_{k}\in\left[0,1\right)
}
}\left\{ \sum_{k}c_{k},\;x=\sum_{k}c_{k}a\left(f_{k},\phi_{k}\right)\right\} .\label{eq:AtomicNormDef}
\end{align}
Atomic norm regularization consists in approaching the pair of sets
$\left(\left\{ f_{k}^{0}\right\} ,\left\{ c_{k}^{0}\right\} \right)$
characterizing the distribution $\hat{x}_{0}$ with the pair $\left(\left\{ f_{k}^{\mathcal{A}}\right\} ,\left\{ c_{k}^{\mathcal{A}}e^{i2\pi\phi_{k}^{\mathcal{A}}}\right\} \right)$
minimizing (\ref{eq:AtomicNormDef}) for a given observation vector
$x$. It was shown in \cite{Bhaskar2013} that this estimate is exact,
under proviso of a minimal spectral separation discussed earlier.
Moreover, it was proven that one has the equality

\begin{align}
\left\Vert x\right\Vert _{\mathcal{A}} & =\inf_{\substack{u\in\mathbb{C}^{n}\\
t>0
}
}\left\{ \frac{1}{2n}\trace\left(\toep_{n}\left(u\right)\right)+\frac{1}{2}t:\;\begin{bmatrix}\mathcal{\toep}_{n}\left(u\right) & x\\
x^{\herm} & t
\end{bmatrix}\succeq0\right\} ,\label{eq:AtomicNormSDP}
\end{align}
where the infimum is reached over a pair $\left\{ u_{\mathcal{A}},t_{\mathcal{A}}\right\} $
for which the positive Toeplitz matrix $\mathcal{\toep}_{n}\left(u_{\mathcal{A}}\right)$
admits for eigenvectors the atoms $\left\{ a\left(f_{k}^{\mathcal{A}},0\right)\right\} $.
Hence, Relation (\ref{eq:AtomicNormSDP}) provides a direct way to
recover the spectral support of $\hat{x}_{0}$ by solving a \emph{convex}
semidefinite program (SDP) of dimension $n+1$.

\section{Partial line spectral estimation}

\subsection{Problem statement}

We introduce the partial line spectral estimation problem by extension
of the settings presented in Section \ref{sec:Introduction}. The
sampled vector $x\in\mathbb{C}^{n}$ generated according to Model
(\ref{eq:SpikesModel}) is now assumed to be unknown and one observe
instead linear combinations $y\in\mathbb{C}^{m}$ of $x$ through
a known fat measurement matrix $M\in\mathbb{C}^{m\times n}$, so that
$y=Mx.$

Identically to the original problem, our aim is to recover the sparsest
spectral distribution $\hat{x}_{M,0}\in D_{1}$ matching the measurement
vector $y$, which can be interpreted as the solution of the non-convex
minimization program

\begin{equation}
\hat{x}_{M,0}=\arg\min_{\hat{x}\in D_{1}}\left\Vert \hat{x}\right\Vert _{0},\;\text{subject to }y=M\mathcal{F}_{n}\left(\hat{x}\right).\label{eq:L0-PartialLineSpectralEstimation}
\end{equation}

One could, of course, solve Problem (\ref{eq:L0-PartialLineSpectralEstimation})
by regularizing it on $\mathcal{A}$, generalizing the approach followed
in \cite{Tang2013}; and add the extra linear measurement constraint
$y=Mx$ to Relation (\ref{eq:AtomicNormSDP}), leading to the semidefinite
program
\begin{align}
\left(u_{M,\mathcal{A}},t_{M,\mathcal{A}}\right) & =\arg\min_{\substack{u\in\mathbb{C}^{n}\\
t>0
}
}\frac{1}{2n}\trace\left(\toep_{n}\left(u\right)\right)+\frac{1}{2}t\nonumber \\
\textrm{subject to} & \phantom{=}\begin{bmatrix}\mathcal{\toep}_{n}\left(u\right) & x\\
x^{\herm} & t
\end{bmatrix}\succeq0\nonumber \\
 & \phantom{=}y=Mx.\label{eq:FullSDP}
\end{align}
One can recover in a second time the spectral support via an eigen-decomposition
of $\toep_{n}\left(u_{M,\mathcal{A}}\right)$. However, the SDP (\ref{eq:FullSDP})
involves a cost function and matrix constraint of dimension $n$,
and such approach would require about $\mathcal{O}\left(n^{7}\right)$
operations using standard interior point methods such as \noun{SDPT3}
or \noun{SuDeMi, }whereas the \emph{essential dimension} of the problem
is of order $m$, potentially much smaller than $n$.

\subsection{Contributions}

The rest of this work aims to study two problematics emerging from
the statement of the non-convex Program (\ref{eq:L0-PartialLineSpectralEstimation}). 
\begin{itemize}
\item How to \emph{efficiently} solve Program (\ref{eq:L0-PartialLineSpectralEstimation})
via atomic norm minimization? And, in particular, is there a semidefinite
representation of dimension $m$ for this problem?
\item Can we find sub-sampling matrices $M$ for which \emph{exact recovery}
of the spectral support of $\hat{x}_{0}$ is possible from the sole
observation of the vector $y$?
\end{itemize}
The first question is answered in Section \ref{sec:Main-results},
where we derive, after proving a novel extension of the Carathéodory-Toeplitz
lemma, conditions on the sub-sampling matrix $M$ under which the
partial line spectral estimation problem can be assimilated to an
SDP of dimension $m+1$. In Section \ref{sec:Tightness-of-partial},
we provide theoretical tightness guarantees obtained by generalizing
the Lagrange dual properties studied in \cite{Candes2014a,Duval2015}
onto the partial measurement case. At last, in Section \ref{sec:Applications},
we review certain sub-sampling patterns for which it is possible to
guarantee the recoverability of the spectral support of $\hat{x}^{\star}$
from partial measurements. We illustrate that \emph{poly-logarithmic
time} recovery of $\hat{x}^{\star}$ is possible, and point out the
\emph{sub-Nyquist sampling} capabilities of such approaches.

\section{Main results\label{sec:Main-results}}

\subsection{Partial atomic norm relaxation}

Given a sub-measurement matrix $M\in\mathbb{C}^{m\times n}$ with
$m\leq n$, we define the associated partial atomic set $\mathcal{B}_{M}$
as follows
\[
\mathcal{B}_{M}=M\mathcal{A}=\left\{ b_{M}\left(f,\phi\right),\;f\in\left[0,1\right),\phi\in\left[0,2\pi\right)\right\} ,
\]
where $b_{M}\left(f_{k},\phi_{k}\right)=Ma\left(f_{k},\phi_{k}\right)$.
Due to the absence of ambiguity on the matrix $M$ in this work, the
notations will be simplified to $\mathcal{B}_{M}=\mathcal{B}$ and
$b_{M}\left(f_{k},\phi_{k}\right)=b\left(f_{k},\phi_{k}\right)$.
Similarly to Equations (\ref{eq:AtomicNormGauge}) and (\ref{eq:AtomicNormDef}),
the partial atomic norm $\left\Vert \cdot\right\Vert _{\mathcal{B}}$
is defined for all $y\in\mathbb{C}^{m}$ by the gauge of $\mathcal{B}$,
i.e. 
\begin{align}
\left\Vert y\right\Vert _{\mathcal{B}} & =\inf_{t>0}\left\{ x\in t\,{\rm conv}\left(\mathcal{B}\right)\right\} \nonumber \\
 & =\inf_{\substack{c_{k}\geq0,\;\phi_{k}\in\left[0,2\pi\right)\\
f_{k}\in\left[0,1\right)
}
}\left\{ \sum_{k}c_{k},\;y=\sum_{k}c_{k}b\left(f_{k},\phi_{k}\right)\right\} .\label{eq:PartialAtomicNormDef}
\end{align}
Once again the partial atomic ball verifies the property of being
the smallest convex body containing all the building blocks $b\left(f,\phi\right)$.
Moreover, the partial atomic norm relaxation consists in estimating
the spectral support of $\hat{x}_{M,0}$ with the triplet of sets
$\left(\left\{ f_{k}^{\mathcal{B}}\right\} ,\left\{ c_{k}^{\mathcal{B}}\right\} ,\left\{ \phi_{k}^{\mathcal{B}}\right\} \right)$
realizing the infimum of (\ref{eq:PartialAtomicNormDef}).

\subsection{Semidefinite representability of partial atomic norms}

The theoretical description (\ref{eq:PartialAtomicNormDef}) of partial
atomic norm minimization does not provide an efficient way to compute
an atomic decomposition on $\mathcal{B}$. On the other hand, the
semidefinite representability of the atomic decomposition on $\mathcal{A}$
provided in Equation (\ref{eq:AtomicNormSDP}) holds very specifically
in the line spectral estimation framework due to the close link between
complex exponential vectors of $\mathbb{C}^{n}$ and Toeplitz matrices
of same dimension. It was demonstrated by Carathéodory in \cite{caratheodory1911variabilitatsbereich}
 that the eigenvectors of positive Hermitian Toeplitz matrices are
always elements of the form $a\left(f,0\right)$.

In this section we propose in Lemma \ref{lem:CaratheodoryLemma} a
novel extension of Carathéodory's result, establishing a relationship
between the elements of the form $b\left(f,0\right)\in\mathcal{B}$
and the linear operator $\rmop$. We latter conclude in Theorem \ref{thm:SemidefiniteRepresentability}
on the semidefinite representability of the norm $\left\Vert \cdot\right\Vert _{\mathcal{B}}$.
\begin{lem}
[Partial Carathéodory-Toeplitz lemma]\label{lem:CaratheodoryLemma}

Consider a matrix $M\in\mathbb{C}^{m\times n}$, $m\leq n$, satisfying
the following properties:
\begin{itemize}
\item $M$ is full rank.
\item $\mathcal{T}_{n}^{\herm}\left(M^{*}M\right)$ belongs to the range
of $M^{\herm}$
\end{itemize}
then any positive matrix $S$ of rank $r$ in the range of the operator
$\rmop$ can be decomposed under the form
\[
S=VDV^{\herm},
\]
where
\begin{align*}
V & =\left[b\left(f_{1},0\right),\cdots,b\left(f_{r},0\right)\right]\\
D & ={\rm diag}\left(\left[d_{1},\cdots,d_{r}\right]\right),
\end{align*}
and whereby $d_{k}>0$ are all positive real numbers.

\end{lem}
\begin{IEEEproof}
Every vector $u\in\mathbb{C}^{n}$ can be decomposed under the form
\[
u=w+z,
\]
where $w\in\textrm{range}\left(M^{\herm}\right)$ and $z\in\textrm{range}\left(M^{\herm}\right)^{\perp}=\textrm{ker}\left(M\right)$.
Writing $w=M^{\herm}v$, it comes by linearity, for any $u\in\mathbb{C}^{n}$
\[
\toep\left(u\right)=\toep\left(M^{\herm}v\right)+\toep\left(z\right)
\]
and
\begin{align}
M\toep\left(u\right)M^{\herm} & =M\toep\left(M^{\herm}v\right)M^{\herm}+M\toep\left(z\right)M^{\herm}\nonumber \\
 & =\rmop\left(v\right)+M\toep\left(z\right)M^{\herm}.\label{eq:R-Decomposition}
\end{align}
Moreover, using the properties of the matrix $M$, one has
\begin{align}
M\toep\left(z\right)M^{\herm} & =\left\langle M^{\herm},\toep\left(z\right)M^{\herm}\right\rangle \nonumber \\
 & =\left\langle \toep^{\herm}\left(M^{\herm}M\right),z\right\rangle .\label{eq:R-inner product}
\end{align}
Since $\toep^{\herm}\left(M^{\herm}M\right)$ belongs by assumption
to the range of the matrix $M^{\herm}$, its inner product with the
vector $z$ is null. Consequently, combining the previous (\ref{eq:R-Decomposition})
and (\ref{eq:R-inner product}), one gets
\begin{align}
M\toep\left(u\right)M^{\herm} & =\rmop\left(v\right).\label{eq:CTlemma-Toeplitz relation}
\end{align}

Now, let $S=\rmop\left(v\right)$ be a positive matrix in the range
of $\rmop$. Using a dimension argument, there must exist at least
one vector $z\in{\rm range}\left(M^{\herm}\right)^{\perp}$ for which
the completion $\toep\left(u\right)=\toep\left(M^{\herm}v\right)+\toep\left(z\right)$
is a positive matrix. By application of the Carathéodory-Toeplitz
lemma \cite{caratheodory1911variabilitatsbereich}, there exist a
matrix $U=$$\left[a\left(f_{1},0\right),\cdots,a\left(f_{r},0\right)\right]$
and a positive diagonal matrix $D={\rm diag}\left(\left[d_{1},\cdots,d_{r}\right]\right)$
such that
\[
\toep\left(u\right)=UDU^{\herm}.
\]
One concludes using Equation (\ref{eq:CTlemma-Toeplitz relation})
that
\begin{align*}
\rmop\left(v\right) & =MUDU^{\herm}M^{\herm}\\
 & =VDV^{\herm},
\end{align*}
by letting $V=MU$, which concludes the proof.
\end{IEEEproof}
We are know ready to state the main contribution of this work. The
proof structure of this result is close to the one presented in \cite[Proposition II.1]{Tang2013}
for the case of fully observed atomic norm minimization problems.
\begin{thm}
[Semidefinite representability of partial atomic sets]\label{thm:SemidefiniteRepresentability}Suppose
that $M\in\mathbb{C}^{m\times n}$ satisfies the conditions of Lemma
\ref{lem:CaratheodoryLemma} and that $\sqrt{n}M$ is a unitary matrix,
then for any vector $y\in\mathbb{C}^{m}$, the following equality
holds
\begin{align}
\left\Vert y\right\Vert _{\mathcal{B}} & =\inf_{v\in\mathbb{C}^{m}}\left\{ \frac{1}{2m}\trace\left(\rmop\right)+\frac{1}{2}t,\quad\begin{bmatrix}\rmop\left(v\right) & y\\
y^{\herm} & t
\end{bmatrix}\succeq0\right\} .\label{eq:PartialAtomicSDPEquivalence}
\end{align}
\end{thm}
\begin{IEEEproof}
First of all, since $\sqrt{n}M$ is unitary, one has $\left\Vert b\left(f,\phi\right)\right\Vert _{2}^{2}=\trace\left(Ma\left(f,\phi\right)a\left(f,\phi\right)^{\herm}M^{\herm}\right)=\trace\left(I_{m}\right)=m$
for all $f\in\left[0,1\right)$ and $\phi\in\left[0,2\pi\right)$.

Denote by ${\rm SDP}\left(y\right)$ the quantity on the right hand
side of (\ref{eq:PartialAtomicSDPEquivalence}). Suppose a decomposition
of $y\in\mathbb{C}^{m}$ under the form $y=\sum_{k}c_{k}b\left(f_{k},\phi_{k}\right)$
with $c_{k}>0$, and denote by $v\in\mathbb{C}^{m}$ the phaseless
counterpart of $y$ given by $v=\sum_{k}c_{k}b\left(f_{k},0\right)$.
Moreover, let $t=\sum_{k}c_{k}$ and $u=\sum_{k}c_{k}a\left(f_{k},0\right)$.
Using (\ref{eq:CTlemma-Toeplitz relation})
\begin{align}
\rmop\left(v\right) & =M\sum_{k}c_{k}a\left(f_{k},0\right)a\left(f_{k},0\right)^{\herm}M^{\herm}\nonumber \\
 & =M\sum_{k}c_{k}a\left(f_{k},\phi_{k}\right)a\left(f_{k},\phi_{k}\right)^{\herm}M^{\herm}\nonumber \\
 & =\sum_{k}c_{k}b\left(f_{k},\phi_{k}\right)b\left(f_{k},\phi_{k}\right)^{\herm}.\label{eq:PhaselessRelationship}
\end{align}
Thus, the matrix form of interest can be identified as follows
\[
\begin{bmatrix}\rmop\left(v\right) & y\\
y^{\herm} & t
\end{bmatrix}=\sum_{k}c_{k}\begin{bmatrix}b\left(f_{k},\phi_{k}\right)\\
1
\end{bmatrix}\begin{bmatrix}b\left(f_{k},\phi_{k}\right)\\
1
\end{bmatrix}^{\herm},
\]
and is therefore a positive Hermitian matrix. Equation (\ref{eq:PhaselessRelationship})
ensures that $\trace\left(\rmop\left(v\right)\right)=m\sum_{k}c_{k}$,
and therefore that ${\rm SDP}\left(y\right)\leq\sum_{k}c_{k}$. Consequently
one has $\left\Vert y\right\Vert _{\mathcal{B}}\geq{\rm SDP}\left(y\right)$.

It remains to show $\left\Vert y\right\Vert _{\mathcal{B}}\leq{\rm SDP}\left(y\right)$
to conclude the proof. Suppose that there exist some $v\in\mathbb{C}^{m}$
such that
\begin{equation}
\begin{bmatrix}\rmop\left(v\right) & y\\
y^{\herm} & t
\end{bmatrix}\succeq0,\label{eq:SDP-relationship}
\end{equation}
which implies that $\rmop\left(v\right)\succeq0$. By application
of Lemma \ref{lem:CaratheodoryLemma}, it is possible to decompose
$\rmop\left(v\right)$ under the form
\[
\rmop\left(v\right)=VDV^{\herm}=\sum_{k}d_{k}b\left(f_{k},0\right)b\left(f_{k},0\right)^{\herm},
\]
and the relationship $\trace\left(\rmop\left(v\right)\right)=m\trace\left(D\right)$
holds. Observing that (\ref{eq:SDP-relationship}) implies that $y$
lies in the range of $V$, there exists a vector $w\in\mathbb{C}^{m}$
such that $y=\sum_{k}w_{k}b\left(f_{k},0\right)=Vw.$ Using the Schur
complement lemma, it holds that
\[
VDV^{\herm}\succeq\frac{1}{t}yy^{\herm}=\frac{1}{t}Vww^{\herm}V^{\herm}.
\]
Call $q\in\mathbb{C}^{m}$ the vector solution of $V^{\herm}q={\rm sign}\left(w\right)$,
which exists since $V^{\herm}$ is full rank. We have that
\begin{align*}
\trace\left(D\right) & =q^{\herm}VDV^{\herm}q\\
 & \succeq\frac{1}{t}q^{\herm}Vww^{\herm}V^{\herm}q\\
 & =\frac{1}{t}\left(\sum_{k}\left|w_{k}\right|\right)^{2}.
\end{align*}
This implies using the geometric mean comparison lemma that
\begin{align*}
\frac{1}{2m}\trace\left(\rmop\left(v\right)\right)+\frac{1}{2}t & =\frac{1}{2}\trace\left(D\right)+\frac{1}{2}t\\
 & \geq\sqrt{\trace\left(D\right)t}\\
 & \geq\sum\left|w_{k}\right|\geq\left\Vert y\right\Vert _{\mathcal{B}},
\end{align*}
which completes the proof.
\end{IEEEproof}
\begin{rem}
\label{rem:SelectionMatrix}Lemma \ref{lem:CaratheodoryLemma} and
Theorem \ref{thm:SemidefiniteRepresentability} both require the sub-sampling
$M$ to satisfy a bilinear relationship of the form $\toep_{n}^{\herm}\left(M^{\herm}M\right)-M^{\herm}v=0$
for some $v\in\mathbb{C}^{m}$. Although it is challenging to explicit
the set of matrices satisfying this property, it is trivial to verify
this hypothesis for a given matrix $M$. Moreover, many practical
sub-sampling patterns do satisfy this relation. For instance, consider
a selection matrix $M=C_{\mathcal{I}}$ whose rows are equal to $\left\{ e_{j},\,j\in\mathcal{I}\right\} $
for some subset $\mathcal{I}\subseteq\left\llbracket 0,n-1\right\rrbracket $
for cardinality $m$. This category of sub-sampling matrices corresponds
to practical signal processing sampling schemes where the sample $x_{j}$
is either kept, if $j\in\mathcal{I}$, either discarded. It can be
easily verified that 
\begin{align*}
\mathcal{T}_{n}^{\herm}\left(C_{\mathcal{I}}^{*}C_{\mathcal{I}}\right) & =\mathcal{T}_{n}^{\herm}\left({\rm diag}\left(\mathcal{I}\right)\right)\\
 & =me_{0},
\end{align*}
whereby $e_{0}\in\mathbb{C}^{n}$ is the first vector of the canonical
basis. Therefore, it comes that $C_{\mathcal{I}}$ satisfies the desired
properties if and only if $0\in\mathcal{I}$. Examples involving such
sub-sampling matrices will be discussed in more details in Section
\ref{sec:Applications}.
\end{rem}
Finally, combining the definition of the partial atomic relaxation
(\ref{eq:PartialAtomicNormDef}) with Theorem \ref{thm:SemidefiniteRepresentability}
leads to the following corollary.
\begin{cor}
If the sub-sampling matrix $M\in\mathbb{C}^{m\times n}$ satisfies
the conditions of Theorem \ref{thm:SemidefiniteRepresentability},
then the estimate $\hat{x}_{\mathcal{B}}$ of the spectral support
of $\hat{x}_{0}$ can be computed by solving the semidefinite program

\begin{align}
\left(v_{\mathcal{B}},t_{\mathcal{B}}\right) & =\arg\min_{\substack{v\in\mathbb{C}^{m}\\
t>0
}
}\frac{1}{2m}\trace\left(\rmop\left(v\right)\right)+\frac{1}{2}t\nonumber \\
\textrm{subject to} & \phantom{=}\begin{bmatrix}\rmop\left(v\right) & y\\
y^{\herm} & t
\end{bmatrix}\succeq0,\label{eq:ReducedSDP}
\end{align}
whereby the estimated supporting frequencies $\left\{ f_{k}^{\mathcal{\mathcal{B}}}\right\} $
can be recovered from the vectors $\left\{ b\left(f_{k}^{\mathcal{B}},0\right)\right\} $
forming the columns of $V\in\mathbb{C}^{m\times r}$ where $\rmop\left(v_{\mathcal{B}}\right)=VDV^{\herm}$.
Moreover, the equality $Mu_{\mathcal{A},M}=v_{\mathcal{B}}$ holds
between the outputs Program (\ref{eq:FullSDP}) and Program (\ref{eq:ReducedSDP}).
\end{cor}
The semidefinite program (\ref{eq:ReducedSDP}) is of dimension $m+1$,
and can be solved in polynomial time with respect to $m$ using appropriate
out of the box convex solvers or the Alternating Direction Method
of Multipliers.

\section{Tightness of partial atomic relaxation\label{sec:Tightness-of-partial}}

The guarantees provided in \cite{Candes2014a,Fernandez-granda2015,Duval2015}
for the full line spectral estimation problem are based on the existence
of a polynomial obeying certain extremal properties. Such polynomial
is often refereed as \emph{dual certificate} for the Program (\ref{eq:L0-FullObservation}).
Its existence suffices to guarantee to tightness of the atomic relaxation
as well as the uniqueness of the solution. The next proposition extends
this theory to the partial observation case for an arbitrary matrix
$M\in\mathbb{C}^{m\times n}$. 
\begin{prop}
[Dual certifiability]\label{prop:DualCertifiability}If there exists
a polynomial $Q_{\star}\in\mathbb{C}^{n-1}\left[X\right]$ having
for coefficients vector $q_{\star}\in\mathbb{C}^{n}$ satisfying the
conditions
\begin{equation}
\begin{cases}
q_{\star}\in{\rm range}\left(M^{\herm}\right)\\
Q_{\star}\left(e^{i2\pi f_{k}}\right)={\rm sign}\left(c_{k}\right), & \forall k\in\left\llbracket 1,s\right\rrbracket \\
\left|Q_{\star}\left(e^{i2\pi f}\right)\right|<1, & \text{otherwise},
\end{cases}\label{eq:DualCertificateCondition}
\end{equation}
then the solution of the Program (\ref{eq:L0-FullObservation}) and
the reconstructed estimate $\hat{x}_{\mathcal{B}}$ obtained by solving
(\ref{eq:ReducedSDP}) are unique and verify $\hat{x}_{0}=\hat{x}_{\mathcal{B}}$.
\end{prop}
This proposition is an immediate consequence of the imbrication of
the dual feasible set of Program (\ref{eq:ReducedSDP}) in the one
Program (\ref{eq:L0-FullObservation}), the interested reader is invited
to refer to \cite{FerreiraDaCosta} for the proof details.

Finding explicit sufficient conditions for the existence of such dual
certificate is a difficult problem in the general case. One might
expect their existence to be related to to the separability condition
discussed in Section \ref{sec:Introduction}. Although explicit criterion
have been provided for specific categories of matrices, the problem
for arbitrary matrices $M\in\mathbb{C}^{m\times n}$ remains an active
area of research.

\section{Applications\label{sec:Applications}}

In this last section, we discuss the benefits of the novel semidefinite
formulation provided in Theorem \ref{thm:SemidefiniteRepresentability}
for two different categories of sub-sampling patterns $M$. Both of
those sub-sampling patterns fall into the category of selection matrices
$C_{\mathcal{I}}$ introduced in Remark \ref{rem:SelectionMatrix},
and therefore satisfies the conditions of Theorem \ref{thm:SemidefiniteRepresentability}.
We explicit the advantages of the reduced SDP formulation (\ref{eq:ReducedSDP})
in each of those settings.

\subsection{Random sub-sampling}

Random sub-sampling was introduced in the original work \cite{Tang2013}
and is characterized as follows. The observed vector $y$ is constructed
by keeping uniformly at random each of the entries of the sampling
vector $x\in\mathbb{C}^{n}$ independently from the others. The $j^{th}$
observation is inserted with probability $p$ in the vector $y$ and
discarded with probability $1-p$. Supposing that there remain $m$
elements at the end of the process, it was proven in \cite[Theorem I.1]{Tang2013}
that
\[
m\geq C\max\left\{ \log^{2}\frac{n}{\delta},s\log\frac{s}{\delta}\log\frac{n}{\delta}\right\} 
\]
is enough to ensure with probability at least greater than $1-\delta$
that $\hat{x}_{\text{0}}=\hat{x}_{M,\mathcal{A}}$, provided that
$\Delta_{\mathbb{T}}\left(\hat{x}\right)>\frac{4}{n-1}$.
\begin{cor}
Suppose that $\Delta_{\mathbb{T}}\left(\hat{x}\right)>\frac{4}{n-1}$
and suppose that the measurements $y\in\mathbb{C}^{m}$ have been
acquired through a random subsampling process, then the semidefinite
program (\ref{eq:ReducedSDP}) of dimension $m+1=\mathcal{O}\left(\max\left\{ \log^{2}\frac{n}{\delta},s\log\frac{s}{\delta}\log\frac{n}{\delta}\right\} \right)$
returns the optimal line spectrum $\hat{x}_{0}$ of Problem (\ref{eq:L0-FullObservation})
with probability at least greater than $1-\delta$.
\end{cor}
The original formulation of the problem being of dimension $n+1$,
this results bring \emph{order of magnitude }changes to the computational
complexity of line spectral estimation problem. Indeed, the dimension
$m+1$ of SDP (\ref{eq:ReducedSDP}) is \emph{poly-logarithmic} on
the number of initial samples $n$.

\subsection{Multirate sampling systems}

Multirate sampling systems (MRSS) have been studied in \cite{FerreiraDaCosta,FerreiraDaCosta2016}
as a way to estimate sparse spectra in distributed environments. Those
systems are formed by a set of $p$ uniform samplers acquiring $\left\{ n_{l}\right\} _{1\leq l\leq p}$
measures at potentially different delays $\left\{ \gamma_{l}\right\} _{1\leq l\leq p}$
and sampling frequencies $\left\{ F_{l}\right\} _{1\leq l\leq p}$.
The observation vector $y\in\mathbb{C}^{m}$ is obtained by merging
the different outputs of all those samplers. It is shown in \cite{FerreiraDaCosta}
that if the samplers obey a common alignment property on a grid of
$n_{\baro}$ elements and if there exists at least one sampler of
index $l_{\star}$ for which

\vspace{-0.2cm}
\begin{equation}
\begin{cases}
\Delta_{\mathbb{T}}\left(\hat{x}\left(\frac{\cdot}{F_{l_{\star}}}\right)\right)\geq\frac{2.52}{n_{j}-1}\\
n_{j}>2\times10^{3},
\end{cases}\label{eq:MRSS-condition}
\end{equation}
whereby $\hat{x}\left(\frac{\cdot}{F_{l_{\star}}}\right)$ denotes
the normalized spectrum for the sampling frequency $F_{l_{\star}}$,
then the atomic norm relaxation of the line spectral estimation problem
is tight. Since $m\ll n_{\baro}$ up to logarithmic order, Program
(\ref{eq:ReducedSDP}) is particularly efficient in this context.
In addition, it is shown that MRSS provides an efficient way to recover
spectra at sub-Nyquist sampling frequencies. Applying Theorem \ref{thm:SemidefiniteRepresentability}
in this context gives the following corollary.
\begin{cor}
Consider a MRSS verifying the conditions (\ref{eq:MRSS-condition}),
then the semidefinite program (\ref{eq:ReducedSDP}) of dimension
$m+1\ll n_{\baro}$ returns the optimal solution of line spectral
estimation problem (\ref{eq:L0-FullObservation}). Moreover, the ground
truth spectrum $\hat{x}_{\star}$ can be reconstructed at sub-Nyquist
rates up to a spectral aliasing factor modulo $F_{\baro}\sim\prod_{l=1}^{p}F_{l}$.
\end{cor}
\bibliographystyle{IEEEtran}
\bibliography{BibTeX}

\end{document}